\documentclass[structabstract]{aa}  
\usepackage{graphicx}
\usepackage{txfonts}
\usepackage{multirow}
\usepackage{natbib}
\bibpunct{(}{)}{;}{a}{}{,}

\begin{document}

   \title{X-ray stacking of Lyman break galaxies in the 4\,Ms CDF-S}

   \subtitle{X-ray luminosities and star formation rates across cosmic time}

   \author{Peter-Christian Zinn\inst{1,2}
          \and
          Stefan Blex\inst{1}
          \and
          Nicholas Seymour\inst{2}
          \and
          Dominik J. Bomans\inst{1}
          }

   \institute{Astronomisches Institut Ruhr-Universit\"at Bochum, Universit\"atsstr. 150, 44801 Bochum, Germany\\
              \email{zinn@astro.rub.de}\\
              CSIRO Astronomy \& Space Science, PO Box 76, Epping, NSW, 1710, Australia}

   \date{Received 25 June 2012; accepted 3 October 2012}

 
\abstract{Lyman Break Galaxies (LBGs) are widely thought to be prototypical young galaxies in the early universe, particularly representative of those undergoing massive events of star formation. Therefore, LBGs should produce significant amounts of X-ray emission.}{We aim to trace the X-ray luminosity of Lyman Break Galaxies across cosmic time and from that derive constraints on their star formation history.}{We utilize the newly released 4\,Ms mosaic obtained with the {\it Chandra} {\rm X-ray Observatory, the deepest X-ray image to date, alongside with the superb spectroscopic data sets available in the CDF-S survey region to construct large but nearly uncontaminated samples of LBGs across a wide range of redshift ($0.5<z<4.5$) which can be used as input samples for stacking experiments. This approach allows us to trace the X-ray emission of Lyman Break Galaxies to even lower, previously unreachable, flux density limits ($\sim10^{-18}\,$mW\,m$^{-2}$) and therefore to larger redshifts.}}{{\rm We reliably detect soft-band X-ray emission from all our input redshift bins except for the highest redshift ($z\sim4$) one. From that we derive rest-frame 2-10\,keV luminosities and infer star formation rates and stellar masses. We find that star formation in LBGs peaks at a redshift of $z_{peak}\approx3.5$ and then decreases quickly. We also see a characteristic peak in the specific star formation rate (sSFR=SFR/M$_*$) at this redshift. Furthermore, we calculate the contribution of LBGs to the total cosmic star formation rate density (SFRD) and find that the contribution of LBGs is negligible. Therefore, we conclude that most of the star formation in the early universe takes place in lower luminosity galaxies as suggested by hierarchical structure formation models.}}


   \maketitle
%

\section{Introduction}
\label{intro}
Lyman Break Galaxies (LBGs) are largely considered as the bright end of the distribution of normal star-forming galaxies across a wide range of redshift.They exhibit a significant scatter in terms of mass \citep{Mannucci2009}, with both a pronounced low-mass \citep{Weatherley2003} and high-mass \citep{Barmby2004} fraction among the entire LBG population. Since they are -- thanks to their exceptionally strong star formation activity -- relatively easy to select at various redshift ranges from photometric observations only, they have been playing a key role in galaxy evolution studies all over the last two decades. This is foremost due to the very effective ``drop out'' selection technique established in the early 1990s \citep[see e.g.][]{Steidel1993,Steidel1996,Steidel1999}. This technique utilizes the strong absorption of all light emitted blueward the rest-frame wavelength of the Lyman limit at $912\,\rm{\AA}$. Because this produces a very pronounced step in the typical spectrum of an LBG, one can select such objects by searching deep imaging data for sources detected only in longer wavelength filters but not in short wavelength filters. This behavior of a source ``dropping out'' from being detected below a certain wavelength is nowadays the major tool for selecting candidate high-redshift sources. Since its first application in the early 1990s, the technique underwent a massive development and heavy usage troughout the community \citep[review by][]{Giavalisco2002}. Therefore, large samples of LBGs across the entire redshift range from $z=1$ \citep{Burgarella2007,Haberzettl2009,Basu-Zych2011} to $z>7$ \citep{Bouwens2010,Bouwens2011b,Stark2011} are publicly available, typically comprising hundreds of objects in a comprehensive form. Even at the highest redshifts, considerable work is being done utilizing forefront equipment such as the new Wide-Field Camera 3 (WFC\,3) aboard the {\it Hubble} Space Telescope (HST). Therefore, LBG candidates with redshifts as large as $z=10$ are being discovered \citep{Bouwens2011a}, but because of their extreme faintness ($H_{\mathrm{AB}}=28.9$) lacking spectroscopic confirmation which has only been done for LBGs up to $z\sim7$ \citep{Vanzella2011}. There is also considerable work characterizing LBGs in terms of their environment \citep{Tasker2006,Cook2010}.

Since LBGs are traditionally selected in the optical wavelength regime, recently extending to the near-infrared (NIR) as the WFC\,3 aboard HST became operational, ancillary observations are necessary to characterize these objects over the entire electromagnetic spectrum. This has been done by several groups, successfully detecting individual LBGs at moderate redshifts ($z<3$) also in the mid- and far-infrared (MIR \& FIR) regimes \citep{Rigopoulou2006,Magdis2008,Burgarella2011} thanks to other space-based facilities such as the {\it Spitzer} Space Telescope and, recently, the {\it Herschel} Space Observatory. At even longer wavelength, ground-based observations in the sub-millimeter and radio regimes have been conducted, yielding only a few detections of individual objects at 850\,$\mu$m \citep[e.g.][]{Chapman2009} whereas there is no direct detection of a significantly redshifted LBG to date except for strongly lensed systems such as the ``8 o'clock arc', for instance \citep{Volino2010}.

To access even more redshifted LBGs over the entire spectrum, stacking techniques have been successfully applied by many groups. For instance, the radio properties of LBGs have been discussed with respect to their star formation activity by \cite{Carilli2008}, utilizing the deep Very Large Array (VLA) 1.4\,GHz (or 20\,cm) observations of the COSMOS field. Similar stacking investigations of the star formation history of a sample of very high redshift ($z>7$) LBGs were done by \cite{Labbe2010} in the NIR regime, utilizing ultra-deep {\it Spitzer} data in the {\it Chandra} Deep Field South (CDF-S). This particular survey field is also extremely valuable to study LBGs at the shortest wavelength since it comprises the deepest X-ray observations obtained so far. However, with less deep data significant work on stacking LBGs in the X-ray regime has already been done. A first attempt has been done already \cite{Brandt2001} in the {\it Chandra} Deep Field North (CDF-N) with a 1\,Ms exposure. Stacking a sample of only 24 LBGs with redshifts between $2\le z\le4$, they found a soft-band signal at a significance level of 99.9\%. From that, they calculated an average X-ray luminosity of $3.2\,10^{41}\,\mathrm{erg\,s^{-1}}$ in the rest-frame 2-8\,keV band, comparable to the most luminous local starburst galaxies such as NGC\,3256 \citep{Moran1999}. A similar result was obtained by \cite{Nandra2002} who extended the small sample of \citeauthor{Brandt2001} to a statistically more robust number of 148 LBGs. With this larger sample, they were able to exclude LBGS containing Active Galactic Nuclei (AGN) and hence giving an estimate for the star formation rate of an average LBG at $z\sim3$ of about 60\,$M_{\odot}$\,yr$^{-1}$. A next step was done by \cite {Lehmer2005} utilizing {\it Chandra} data from both the Great Observatories Origins Deep Survey (GOODS) north and south fields with exposure times of 2\,Ms and 1\,Ms, respectively. With a large sample of LBGs comprising a few thousand galaxies, they were able to do their analysis in different redshift bins in the range $3\le z\le6$, detecting a significant signal up to $z=4$. They found the average star formation rate at this redshift decreases to 10-30\,$M_{\odot}$\,yr$^{-1}$, and even further to higher redshifts since they were not able to make statistically significant detections at any redshift greater than $z=4$. A complementary result was obtained by \cite{Laird2006} who identified a sample of direct X-ray detections of LBGs at intermediate redshifts ($z\sim3$), supporting the notion that LBGs are the ``tip of the iceberg'' in terms of star formation in their respective epoch. The statistical X-ray detection of high-redshift LBGs ($z>5$) was attempted for the first time by \cite{Cowie2011} utilizing photometrically selected samples of LBGs comprising several thousand sources in the CDF-S. Although with such a large number of stacked sources, they were not able to detect X-rays from the high-$z$ LBG population, placing an upper limit of $4\,10^{41}\,\mathrm{erg\,s^{-1}}$ on the X-ray luminosity in the rest-frame 4-15\,keV band at $z=6.5$.
 
With this paper, we attempt to widen the redshift range of X-ray studies of LBGs down to $z=1$ since space-based facilities such as GALEX and SWIFT now provide reliable LBG samples selected in the ultraviolet (UV). Furthermore, we want to trace the star formation activity and stellar mass build-up of the average LBG from $z=1$ onwards.
 
Throughout this paper, we adopt a standard flat $\Lambda$CDM cosmology with $H_{\mathrm{0}}=70\,\mathrm{km\,s^{-1}\,Mpc^{-1}}$ and $\Omega_{\mathrm{\Lambda}}=0.73$ \citep{Komatsu2011}.

\section{X-ray data and LBG selection}
In this paper, we use the 4\,Ms {\it Chandra} Deep Field South data\footnote{\texttt{http://www2.astro.psu.edu/users/niel/\\cdfs/cdfs-chandra.html}} as basis for the X-ray stacking procedure. These data comprise mosaics of an area of 464.5\,arcmin$^{2}$ in both the soft (0.5-2\,keV) and the hard (2-8\,keV) X-ray bands. With an effective exposure time of 3822\,ks, this is the the deepest X-ray data set available to date, reaching an on-axis flux limit of $9.1\,10^{-18}\,\mathrm{erg\,s^{-1}\,cm^{-2}}$ and $5.5\,10^{-17}\,\mathrm{erg\,s^{-1}\,cm^{-2}}$ in the soft and hard band \citep{Xue2011}.

To create a parent list of Lyman Break Galaxies to be stacked, we use the extreme wealth of spectroscopic data available in the CDF-S. To use only spectroscopically confirmed LBGs assures both a clean list with only marginal contamination by mis-classified objects, e.g. lower redshift interlopers, as well as accurate redshifts for the construction of several redshift bins. In particular, we used the VLT/VIMOS spectroscopic surveys \citep{Popesso2009} and \cite{Balestra2010} as well as the VLT/FORS survey by\cite{Vanzella2005,Vanzella2006,Vanzella2008}. These catalogs provide a detailed spectroscopic classification and are therefore ideal for reliably selecting LBGs. The main photometric selection criterion was based on a $U$- and $B$-band drop-out search utilizing a color cut $U-B\ge1.2$ or $B-V\ge1.2$, respectively, where the red part of the spectrum must stay flat (e.g. $V-z\le1.2$). The main spectral feature used in these surveys for deeming a source an LBG is the characteristic break in their spectrum blueward of the Lyman Limit at rest-frame $912\,\rm{\AA}$ which was assessed by comparison to various template spectra of Lyman Break Galaxies and other, possibly contaminating, source types. The spectral classifications provided by these surveys are very detailed, allowing e.g. for a split in Ly\,$\alpha$ emitters and absorbers. However, in order not to bias our samples in a certain way, we chose to not use this criterion but solely select LBGs (wether they are Ly\,$\alpha$ emitters or absorbers) in different redshift bins. For a complete overview of our input list see Tab.~\ref{table:1}.

\begin{table*}
\caption{LBG selection parent tables and redshift bins.}
\label{table:1}
\centering
\begin{tabular}{c c r r r r r}
\hline
\hline
\noalign{\smallskip}
ID & parent list & redshift range & median $z$ & age\tablefootmark{a} [Gyr] & \# selected objects& \# stacked objects \\\hline
ref-lowZ & \cite{Wuyts2008} & $0.5\leq z\le 1.0$ & 0.68 & 7.52 & 143 & 141 \\
BZ-UV & \cite{Basu-Zych2011} & $1.0\leq z\le 2.0$ & 1.33 & 4.82 & 43 & 35 \\
Spec-U1 & VIMOS \& FORS\tablefootmark{b} & $2.5\leq z\le 3.0$ & 2.67 & 2.49 & 201 & 201 \\
Spec-U2 & VIMOS \& FORS\tablefootmark{b} & $3.0\leq z\le 3.5$ & 3.33 & 1.95 & 143 & 92 \\
Spec-B1 & VIMOS \& FORS\tablefootmark{b} & $3.5\leq z\le 4.0$ & 3.70 & 1.73 & 63 & 19 \\
Spec-B2 & VIMOS \& FORS\tablefootmark{b} & $4.0\leq z\le 4.5$ & 4.16 & 1.50 & 39 & 34 \\
\hline
\end{tabular}
\tablefoot{
    \tablefoottext{a}{Time since the Big Bang according to the cosmology defined in Sect.~\ref{intro} evaluated at median redshift of the bin.}
  \tablefoottext{b}{\cite{Popesso2009}, \cite{Balestra2010}, \cite{Vanzella2008}.}}
\end{table*}

To extend our LBG sample to redshifts lower than $z=2.5$, the approximate limit for the spectroscopic classification of LBGs using optical spectra, we had to take a closer look into the UV data of the CDF-S because at these low redshifts, the Lyman break occurs in the near UV range. Using the well-known dropout technique to identify Lyman Break galaxy candidates, \cite{Basu-Zych2011} compiled a list of candidates with data from the SWIFT Gamma-ray observatory's UV/optical telescope (SWIFT-UVOT). The SWIFT satellite, although dedicated to high energy astrophysics, is equipped with a small (30\,cm diameter) telescope sensitive to wavelength between 170\,$\mu$m and 650\,$\mu$m which observed an area of 266\,arcmin$^2$ in the CDF-S for approximately 60\,ks, reaching a limiting magnitude in the $U$-band of 24.5\,AB-mag. These observations lead to a list of candidate Lyman Break Galaxies with 43 objects, spanning the desired redshift range. This sample, dubbed BZ-UV, is also listed in Tab.~\ref{table:1}. Because their selection mechanism is based only on photometric data, we point out that the reliability of these objects is not comparable to the spectroscopic lists we use for higher redshift objects. Therefore, and because there are much less LBG candidates identified via UV observations, we expect our stacking results in this redshift range to be significantly less robust than in the other redshift bins.

{ We have taken special care in order to avoid any contamination by X-ray emission from a faint AGN that may be hosted by our stacked sources. The rejection of possible AGN hosts was done in four stages: (i) The optical spectra of all sources in the input lists were examined both by eye and according to the classical BPT diagnostic diagram for AGN activity \citep{Baldwin1981}. Any source with a marginal sign for AGN activity was discarded from the list. (ii) We used the deep {\it Spitzer} photometry in the CDF-S (GOODS project, see \texttt{http://irsa.ipac.caltech.edu/data/GOODS/}) to examine our input LBGs in the mid-infrared color-color diagram according to the criteria from \cite{Stern2005} and excluded all sources lying in the AGN regime of this diagnostic diagram. (iii) We cross-correlated our input source lists to the Very Large Array (VLA) 20\,cm observations by \cite{Miller2008} and plotted them against the {\it Spitzer}/MIPS $24\,\mu$m flux densities to obtain a measure for the radio--infrared correlation amongst our sources. Such sources that deviate from this correlation as defined in \cite{Mao2011} by showing a more than a 5-fold excess in radio emission most likely originating from an AGN were discarded.}

In order to compare our findings for star formation rates which will be calculated from the stacked X-ray luminosities, we use data of our selected objects from other survey projects in the CDF-S. Primarily, the COMBO-17 survey \citep{Wolf2003} is used for its broad range of data available over a large area. A deeper but more narrow view (in terms of covered area) is introduced by the use of data from the GEMS catalog \citep{Rix2004} with {\it Hubble} Space Telescope (HST) observations in two filters and the FIREWORKS \citep{Wuyts2008} catalog, containing deep data across a wide range of the electromagnetic spectrum (UV to MIR) but for only a fraction of our selected LBGs. In particular, we used the optical and NIR ancillary data to infer parameters such as stellar mass for all objects in our stacks. Furthermore, FIREWORKS was used to construct a low redshift control sample of galaxies. This sample should cover the very low redshift end in our analysis and be consistent of galaxies of all kinds to obtain a completely unbiased view of X-ray emission from typical galaxies to be later compared to our LBG results. To compile this reference sample (dubbed ref-lowZ in Tab.~\ref{table:1}), we only applied a redshift cut of $0.5\leq z\le 1.0$ to all objects classified as galaxy in the COMBO-17 catalog that also have FIREWORKS counterparts with reliable spectroscopic redshifts. This left us with a fairly large sample of more than 200 objects, therefore we expect the stacking to deliver an unambiguous X-ray detection.

\section{The stacking procedure}
Before starting the stacking process for the various source samples defined above, special care was taken to exclude individually detected sources from the samples. Therefore, in a first step, we cross-matched all our sources as summarized in Tab.~\ref{table:1} to the {\it Chandra} 4\,Ms source catalog by \cite{Xue2011} and excluded the 13 sources that have X-ray counterparts in this catalog which are mostly lower-z objects. For all remaining sources, $10\arcsec\times10\arcsec$ (resp. $20\times20$ pixels) cutout images from the 4\,Ms soft-band mosaic centered on the (optical or UV derived) source position were created and inspected by eye whether there is a clear but uncatalogued X-ray counterpart. This was only true for two sources which were then also discarded from the further process.

This left us with six source lists (according to our six redshift bins from Tab.~\ref{table:1}) containing only objects without unambiguously detected X-ray counterparts to start the actual stacking algorithm with.

\subsection{The stacking algorithm}
\label{algorithm}
There has already been some effort in doing X-ray stacking of faint sources as e.g. presented in \cite{Nandra2002}, and also stacking procedures in other wavelength ranges, e.g. radio as described by \cite{Carilli2008}, are known. A recent attempt also using the new 4\,Ms {\it Chandra} CDF-S mosaic is presented in \cite{Cowie2011}. They attempt to trace the X-ray emission of galaxies out to $z=8$ by utilizing high-$z$ galaxy samples compiled from new {\it Hubble} Space Telescope Wide Field Camera 3 (WFC 3) observations of the {\it Hubble} Ultra-Deep Field (HUDF) which is part of the CDF-S and applying a weighted mean stacking algorithm based on various quantities (off-axis angle, aperture radii for flux extraction, exact model of noise in aperture) not related to the sources themselves but to the technical layout of the X-ray observations and the reduction process. This technique may enhance the resultant S/N ratio of the final stacked images, it bears the risk of introducing biases or unconstrained statistical effects just because of its complexity. This is discussed in more detail e.g. in \cite{Lehmer2005} and \cite{Hickox2007}. Therefore, we chose to apply a much simpler and straightforward stacking method that, on the one hand, guarantees to not introduce any biases but for the eventual cost of S/N in the final stacked images. Our stacking routine works as follows.

\begin{figure}
\centering
\includegraphics[width=0.3\textwidth]{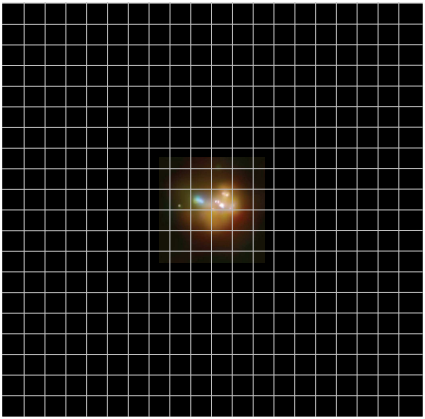}
\caption{The local Lyman Break Galaxy analog VV\,114 as seen by {\it Chandra} \citep[color composite of soft- and hard-band images from][]{Grimes2006}. The overlaid grid indicates the pixel size of the ACIS detector if VV\,114 would be at $z=3$ and our $20\times20$ pixels stack-image layout.}
\label{VV114}
\end{figure}

We first consider the typical angular size of a galaxy at various redshifts to determine a reasonable size for the images to be stacked. Since within our adopted cosmology the linear scale at every redshift $z>1$ is nearly constant with $7\,\mathrm{kpc}/\arcsec$ \citep{Wright2006}, the largest galaxies \citep[determined from their optical morphologies as e.g. by][]{Trujillo2006} at these redshifts with a diameter of $\sim10\,$kpc would stretch over about 1.5$\arcsec$ or three pixels given the pixel scale of {\it Chandra's} ACIS detector. Therefore, we chose to extract fluxes from a 5 pixel diameter aperture to also account for astrometric inconsistencies between the parent galaxy catalogs and the 4\,Ms mosaic. Note that this astrophysical motivation of an aperture size is very much different from other stacking studies such as e.g. \cite{Nandra2002,Lehmer2005}. They defined their aperture sizes empirically by testing various diameters and then adopting the one yielding the best S/N ratio. Despite the fact that those two methods are fundamentally different, the final aperture diamters both our astrophysical and the empirical approach yield are in very good agreement. To finally get a measure of the background around every source, we chose the stacking images to be 20 by 20 pixels, four times the aperture diameter. The layout of our images that go into the actual stacking procedure is shown in Fig.~\ref{VV114}.

Before actually stacking these $20\times20$ pixels images, our algorithm sorts out all images that (i) show an integrated flux within the pre-defined 5 pixel aperture (centered on the image center) that exceeds three times the median absolute deviation (MAD) of the noise in the remaining part of the image and (ii) all images with a total flux across the entire image that is zero. The first step in our algorithm is to reject all marginally detected sources to avoid any contamination of such sources that are neither in the source catalog by \cite{Xue2011} nor visible by eye. This can be due to e.g. AGN which are faint at optical and infrared wavelength and therefore are not visible the FORS or VIMOS spectra \citep[see e.g.][]{Zinn2011}. Since at a redshift $z=3$ the X-ray luminosity necessary to produce an actual detection in the 4\,Ms mosaic is equivalent to a star formation rate of about 200\,$M_{\odot}\,$yr$^{-1}$, an actual X-ray detection is a strong hint for additional AGN activity because LBGs are in general not harboring such extreme starbursts. The second criterion is applied because the 4\,Ms mosaic is smaller than the region for which spectroscopic observations have been carried out. Hence, there may be sources in our list that do not have X-ray coverage which should of course be sorted out in order not to contaminate the final stack. 

To also exclude X-ray detected sources that are just by chance within our stacking region, we define a second aperture of 10 pixels diameter again centered on the image center. The remaining part of the image is used to determine the background noise level. If the total flux measured within this 10 pixel aperture indicates a detection with a significance of more than 90\% with respect to the measured local background, it is rejected for the actual stacking process. This second rejection step is necessary to remove X-ray sources that are close to the stacking location but are not associated with the LBG that should go into the stack. Because those unrelated sources could add flux to the final stack which is not physically related to the LBGs we want to examine, this second rejection step is crucial in terms of a clean stacking sample. Tab.~\ref{table:1} summarizes the number of sources selected for the six input coordinate lists and the number of sources actually stacked after this rejection step.

The actual stacking then consists of a simple average stacking without any weighting in order to not introduce potential biases due to more complex statistical treatments. We point out that, as common in most other wavelength ranges where the underlying noise distribution is Gaussian, we cannot use a median stacking because most of the pixel values stacked are zero, hence the median would always be zero. This is particularly unfortunate because the median is known to be much more robust against outliers (e.g. one unusually bright pixel that immediately increases the average would just be ignored by the median). This is the main reason for applying the rejection as outlined above.

The flux in the six final stacked images is then extracted by integrating over the pre-defined 5 pixel aperture. To maximize the flux covered by the aperture, it is actively centered on the brightest pixel in the image and not fixed at the image center. To correct for noise within the aperture, the background noise is calculated from the remaining part of the image (every pixel not covered by the flux extraction aperture) and then subtracted from the raw flux measured within the aperture. Because not every pixel in the 4\,Ms mosaic has exactly the same expose time, we used the exposure maps distributed with the actual science frames to determine a precise total exposure time for every stack. By dividing the noise-corrected number of counts within our aperture by this total exposure time, we get the final X-ray flux in the respective band in counts per second. The conversion to physical units (here erg\,s$^{-1}$\,cm$^{-2}$=mW\,m$^{-2}$) is then done by extrapolating a conversion factor between counts per second and flux based on the 4\,Ms source catalog by \cite{Xue2011}. To quantify the significance of detection in our final stacked images, the total number of counts within our extraction aperture, dubbed $x$, is compared to the noise level of the stacked image, dubbed $\lambda$. Assuming a Poissonian noise distribution, this noise level should be the expactation value of the distribution. Hence, we calculate a cumulated probability $P$ which is the probability of measuring $x$ or less counts where we expect $\lambda$ counts. If this probability exceeds 95\%, we assume the stacked detection to be real. We point out that, in most cases, our detections are well above the 99\% confidence level. To give an error on our measured flux densities, we compute the number of counts necessary for a 68\% (corresponding to $1\sigma$ confidence level in a Gaussian distribution) signal and attribute this quantity to be the error of our measured flux.

\subsection{Testing the algorithm}
Considering the (mathematical) simplicity of our algorithm, we expect to avoid any biases due to statistical effects. Nevertheless, we ran the following tests on artificial data to ensure the performance of our stacking procedure in terms of reliability of stacked detections. Note that the main assumption of stacking -- the noise goes down with the square-root of the number of stacked objects -- is also valid for Poissonian noise distributions since they also exhibit (on the scales of the stacked sources) uncorrelated noise:

Therefore, we artificially created images with 500 by 500 pixels and a Poissonian noise distribution (as it is the case for X-ray images). Every image contains 27 point sources according to the PSF of {\it Chandra} at random off-axis angels and signal to noise ratios of 3, 1 and 0.1 to obtain nine barely detectable sources, nine not detected sources and nine sources well beyond the detection limit. Because for the {\it Chandra} 4\,Ms mosaic, special care was laid on astrometric calibration \citep[see][section 4.2]{Xue2011} resulting in a median positional offset between X-ray and radio sources from \cite{Miller2008} of $0.24\arcsec$, so clearly below the pixel size of {\it Chandra} which is $0.5\arcsec$, we ignored effects caused positional uncertainties and worked with the exact positions of the artificial sources.

We then stacked the sources according to their attributed S/N ratio using our algorithm as previously described. As usually assumed in stacking procedures, the noise level should decrease with the square root of the number of stacked objects. Therefore we expect the original S/N ratios to be increased by  factor of three, making the barely detected ones well detected, the not detected ones barely detected and the sources well beyond detection should remain undetected. These expectations were met in all three cases: the stack of the nine S/N=3 sources contains one nicely centered source for which we measured a new S/N ratio of 8.8, the nine S/N=1 sources were stacked into a single source with S/N=2.9 and the nine S/N=0.1 sources were not visible in the stacked image at all. The fact that the S/N ratios reached in this test are always a bit below the theoretical expectation of the $\sqrt{N}$-law is mainly just due to small number statistics.

To further verify the algorithm and exclude random effects, e.g. from the fact that there may be a pixel with a high pixel value due to the long tail of the Poissonian noise distribution right at the site of an artificial source and hence unexpectedly increase their S/N ratio, we injected the same 21 sources to the same image again but with a constant offset of five pixels. The sources were then stacked again, resulting into stacked images nearly identical in S/N ratios to the first ones without offsets. For the final test, we stacked the sources with random offsets of up to 5 pixels. As expected, no source could be detected in these stacks (except of course for the nine stacked sources with an initial S/N=3). 

These three tests were repeated on 100 images to assure statistical significance. A summary of the tests is presented in Tab.~\ref{tests}, giving the median S/N ratios of the sources in the final stacked images (in case there is a source detected).
\begin{table}
\caption{Summary of tests of the stacking algorithm. The values given in the table are the median S/N ratios of the final stacked sources over all 100 test runs.}
\label{tests}
\centering
\begin{tabular}{l l|| r r r}
& & \multicolumn{3}{c}{stacking offset} \\
& & none & 5 pixels & random \\
\hline
\hline
\multirow{3}{*}{initial S/N} & 0.1 & nd & nd & nd \\
& 1.0 & 2.91 & 2.88 & nd \\
& 3.0 & 8.88 & 8.89 & nd \\
\end{tabular}
\tablefoot{``nd'' indicates that in the particular stack no source is detected (in the case for a stack of S/N=3 sources no source with a significantly higher S/N).}
\end{table}

\begin{figure*}
\centering
\includegraphics[width=1.0\textwidth]{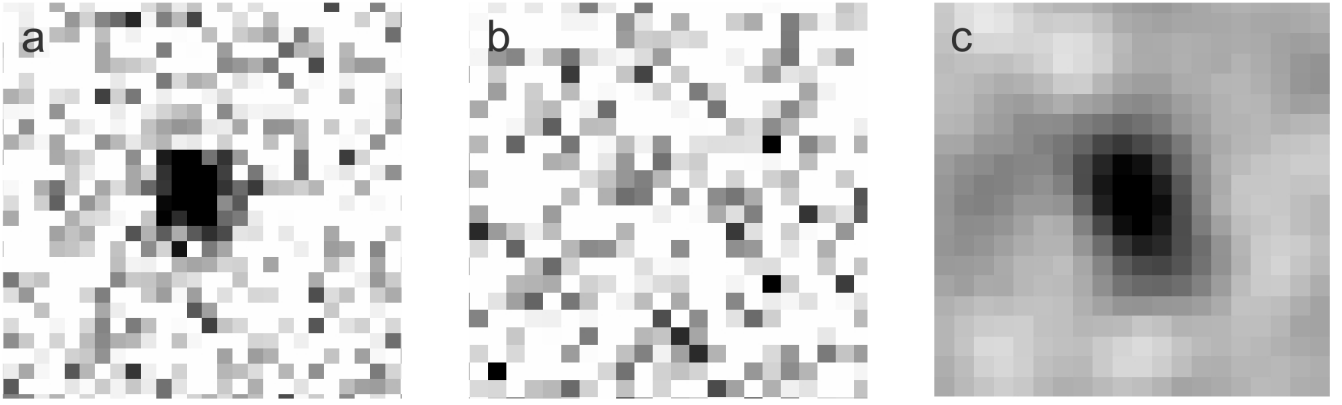}
\caption{Plot illustrating the different steps of our stacking procedure: (a) shows an artificially created source of S/N=5 at the image center within a Poissonian noise distribution, (b) is the same as (a) but with a S/N=1 source and (c) shows the final stacked image of nine such S/N=1 sources (smoothed with a Gaussian of three pixel full-width-half-maximum) using our stacking algorithm.}
\label{testimages}
\end{figure*}

To get a feeling of the stacking as an entire procedure, we also show a panel illustrating our routine (Fig.~\ref{testimages}): The left image shows an artificial source with S/N=5 sitting in a Poissonian noise distribution. The image in the middle is of the same size and noise level but with a source of S/N=1 in the image center and the right image shows the resultant stack of nine such S/N=1 sources (smoothed with a Gaussian of three pixel full-width-half-maximum) using our stacking algorithm. We measured S/N=2.95 for this stack within a circular aperture of 5 pixels diameter. Compare to Fig.~\ref{stacks} showing our stacks with real data.

\section{Stacking results}
\begin{figure*}
\centering
\includegraphics[width=0.95\textwidth]{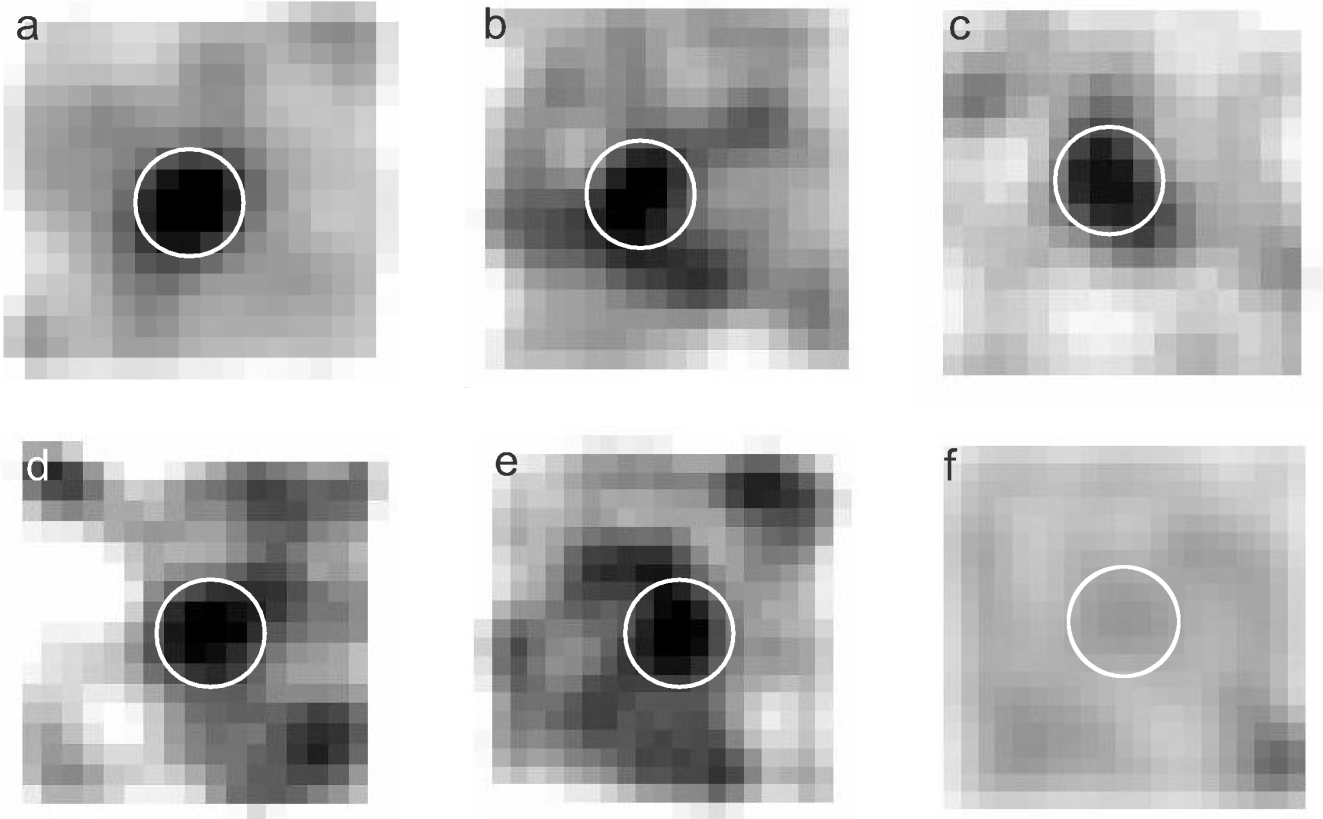}
\caption{The final stacked soft-band images (all smoothed with a Gaussian of FWHM = 3 pixels for illustration purposes) used for further analysis: {\it(a)} ref-lowZ ($P=0.999$), {\it(b)} BZ-UV ($P=0.993$), {\it(c)} Spec-U1 ($P=0.999$), {\it(d)} Spec-U2 ($P=0.958$), {\it(e)} Spec-B1 ($P=0.974$), {\it(f)} Spec-B2 ($P=0.560$). The white circles indicate the 5 pixels diameter apertures used for flux extraction.}
\label{stacks}
\end{figure*}

Our final stacked soft-band images for all six input lists (resp. redshift bins) are shown in Fig.~\ref{stacks}, a table summarizing the extracted fluxes and detection probabilities is presented in Table~\ref{fluxes}.
\begin{table*}
\caption{LBG stacking results.}
\label{fluxes}
\centering
\begin{tabular}{c c r r r r r r r r r}
\hline
\hline
\noalign{\smallskip}
ID & median $z$ & soft & conf\tablefootmark{a} & hard & conf\tablefootmark{a} & $HR$ & soft flux & soft flux err\tablefootmark{b} & hard flux & hard flux err\tablefootmark{b} \\
 & & counts &  & counts & & & 10$^{-18}$\,mW\,m$^{-2}$ & 10$^{-18}$\,mW\,m$^{-2}$ & 10$^{-18}$\,mW\,m$^{-2}$ & 10$^{-18}$\,mW\,m$^{-2}$ \\\hline
ref-lowZ & 0.68 & 153.83 & 0.999 & 60.01 & 0.907 & $<-0.44$ & 2.15 & 0.37 & 4.41 & u.l. \\
BZ-UV & 1.33    & 36.36  & 0.993 & 25.12 & 0.850 & $<-0.18$ & 2.05 & 0.24 & 7.44 & u.l. \\
Spec-U1 & 2.67  & 112.26 & 0.999 & 100.74& 0.971 & $-0.06$ & 1.10 & 0.23 & 5.20 & 1.88 \\
Spec-U2 & 3.33  & 39.91  & 0.958 & 14.54 & 0.645 & $<-0.47$ & 0.85 & 0.19 & 1.64 & u.l. \\
Spec-B1 & 3.70  & 21.52  & 0.974 & 16.83 & 0.722 & $<-0.12$ & 2.23 & 0.69 & 4.19 & u.l. \\
Spec-B2 & 4.16  & 24.14  & 0.660 & 7.09  & 0.626 & ... & 1.40 & u.l. & 2.16 & u.l. \\
\hline
\end{tabular}
\tablefoot{For the convenience of the reader, we note that 1\,mW\,m$^{-2}$=1\,erg\,s$^{-1}$\,cm$^{-2}$.
    \tablefoottext{a}{Probability that this detection is real, assuming a Poissonian noise distribution. See text for further explanation.}
\tablefoottext{b}{u.l. indicates upper limit since no actual detection could be made (confidence $<$ 0.95).}}
\end{table*}

The flux extraction process was performed in two steps: First, all stacks were visually inspected for whether there is a detection or not. To quantify this, we then extracted the number of counts both within our pre-defined 5 pixel aperture (where the aperture was centered on the pixel with the highest count rate for the images where the visual inspection suggested a detection, for the others the aperture was centered on the image center) as well as for the remaining part of the stacked image. From the distribution of counts outside the aperture, we constructed the noise of the stacked image by fitting a Poissonian distribution to the histogram. This was done to verify the Poissonian behavior of the noise in general and to obtain the expectation value, $\lambda$, for every image. From this expectation value, we estimated the noise count rate that would be comprised by our 5 pixel aperture and subtracted it from the actual count rate measured within this aperture (correcting for pixels of which only a fraction lies within the aperture), yielding a count rate of the stacked sources only. Using our fit to the noise distribution, we then calculated the probability that such a count rate occurs just by chance in the respective image. Its inverse probability (so 1 - ``by-chance-probability'') is then attributed the confidence of detection. Adopting a common confidence limit of 95\%, we consider all stacks with a confidence greater than 0.95 as detections. The source count rates (the ones cleaned from noise counts) and the confidence values are given in Table~\ref{fluxes}.

{ We then converted the source count rates to physical flux densities assuming a power-law X-ray spectrum with a photon index typical for star-forming galaxies of $\Gamma$=2.0 as empirically found by \cite{Ptak1999}, see also the discussion on X-ray spectra of LBGs in \cite{Laird2006}. We also corrected for Galactic extinction in the CDF-S ($N_{\mathrm{HI}}=6.8\,10^{19}\,\mathrm{cm}^{-2}$) following \cite{Kalberla2005}.} To test our flux calibration, we compared it to the original 4\,Ms source catalog by \cite{Xue2011} finding them in very good agreement with their fluxes. Errors were assigned to all measured fluxes by taking into account the image noise (more precisely, the 68\% uncertainty of the extracted flux, dubbed $\delta f$) and a possible 10\% error due to the aperture size (dubbed $\sigma$). This error accounts for the fact that we chose our aperture to have a fixed size in contrast to other work where the aperture is empirically determined to maximize the S/N ratio for each stack, see e.g. \cite{Lehmer2005,Laird2006}:
\begin{equation}
\Delta f = \sqrt{(\delta f)^2+(\sigma\cdot f)^2},
\end{equation}
where $f$ is the measured flux of the detected source.

{ To further validate our results regarding the X-ray luminosity computed assuming a power-law with photon index $\Gamma$=2.0, we calculated the hardness ratios (or upper limits, respectively) for each stack and check whether they are consistent with X-ray emission following such a power-law distribution. The hardness ratio is accordingly defined as $HR=(H-S)/(H+S)$, where $H$ and $S$ are the counts in the hard- and soft-band, respectively. Since obscured AGN at the considered redshifts would show rather flat X-ray spectra with photon indices $\Gamma\sim1.0-1.4$, corresponding to $HR>0.0$ \cite{Park2008}, the computed hardness ratios from Table~\ref{fluxes} reveal that also the final stacks are completely consistent with X-ray emission originating from star formation rather than from AGN activity. For completeness, we note that also the soft-band detections and hard-band non-detections for the two low-redshift control samples are consistent with the X-ray luminosity derived from the soft-band fluxes and a power-law SED with $\Gamma=2$.}

\section{Astrophysical interpretation}
Table~\ref{params} summarizes our derived parameters that are discussed in the next sections.
\begin{table*}
\caption{Derived parameters for our LBG samples.}
\label{params}
\centering
\begin{tabular}{c c r r r r r r r r}
\hline
\hline
\noalign{\smallskip}
ID & median $z$ & rest 2-10\,keV\tablefootmark{a} & rest 2-10\,keV err\tablefootmark{b} & SFR & SFR err\tablefootmark{b} & M$_*$ & M$_*$ err & sSFR & sSFR err\tablefootmark{b} \\
 & & $\log(\mathrm{W})$ & $\log(\mathrm{W})$ & $M_{\odot}\,\mathrm{yr}^{-1}$ & $M_{\odot}\,\mathrm{yr}^{-1}$ & $10^{10}\,M_{\odot}$ & $10^{10}\,M_{\odot}$ & Gyr$^{-1}$ & Gyr$^{-1}$ \\\hline
ref-lowZ & 0.68 & 33.06 & 32.30 & 2.3 & 0.40 & 6.73 & 1.61 & 0.03 & 0.01 \\
BZ-UV & 1.33    & 33.65 & 32.72 & 8.93 & 1.05 & 3.35 & 0.67 & 0.27 & 0.08 \\
Spec-U1 & 2.67  & 33.97 & 33.29 & 18.67 & 3.90 & 1.16 & 0.29 & 1.61 & 0.15 \\
Spec-U2 & 3.33  & 34.03 & 33.38 & 21.43 & 4.79 & 1.09 & 0.28 & 1.96 & 0.80 \\
Spec-B1 & 3.70  & 34.53 & 34.02 & 67.77 & 20.97 & 1.09 & 0.21 & 6.25 & 1.49 \\
Spec-B2 & 4.16  & 34.42 & u.l. & 52.61 & u.l. & 1.15 & 0.31 & 4.56 & u.l. \\
\hline
\end{tabular}
\tablefoot{For the convenience of the reader, we note that 1\,W=10$^7$\,erg\,s$^{-1}$.
    \tablefoottext{a}{Assuming a power-law X-ray spectrum with a photon index $\Gamma$=2.0 { typical for LBGs \citep{Laird2006}.} This factor is adopted throughout the stacking literature, see e.g. \cite{Nandra2002} or \cite{Lehmer2005}.}
\tablefoottext{b}{u.l. indicates upper limit since no actual detection could be made (confidence $<$ 0.95).}}
\end{table*}

To obtain the rest-frame 2-10\,keV luminosity, we used our soft-band stacked fluxes and the corresponding median redshift of each bin. The median redshifts were calculated from the original spectroscopic redshifts as given in the input catalogs by \cite{Popesso2009}, \cite{Balestra2010}, and \cite{Vanzella2008} for our spectroscopic samples or, respectively, from the photometric redshifts given in \cite{Basu-Zych2011} for the BZ-UV sample or from the spectroscopic redshifts provided by \cite{Wuyts2008} for the low-z reference sample comprised from the FIREWORKS catalog. To obtain a 2-10\,keV luminosity, we used the  Portable, Interactive Multi-Mission Simulator (PIMMS) version 4.4\footnote{http://asc.harvard.edu/toolkit/pimms.jsp} to get a 2-10\,keV flux based on an extrapolation of our stacked soft-band fluxes and a photon index $\Gamma$=2.0 { typical for LBGs as e.g. found by \cite{Laird2006}}. This factor is adopted throughout the stacking literature, see e.g. \cite{Nandra2002} or \cite{Lehmer2005}. Note that the stacked fluxes from Table~\ref{fluxes} are already corrected for Galactic extinction following \cite{Kalberla2005}. These fluxes could then be directly converted to luminosities with respect to the above specified cosmology. We did not account for any errors due to redshift uncertainties because the flux errors are by far dominating the luminosity errors.

Star formation rates were calculated using the calibration by \cite{Ranalli2003}:
\begin{equation}
SFR\,[M_{\odot}\,\mathrm{yr}^{-1}]=2.0\times10^{-33}L_{2-10\,\mathrm{keV}}
\end{equation}
where $L_{2-10\,\mathrm{keV}}$ is in units of Watts. The corresponding errors take into account both the uncertainties in luminosity and the intrinsic scatter of the calibration. Since \cite{Ranalli2003} did not quantify this scatter, we computed it using their data, finding an rms scatter of roughly 20\%. In addition to the intrinsic scatter in that relation we point out that it is calibrated using a local galaxy sample only and that recent studies \citep[e.g.][]{Dijkstra2012} consider the possibility that the conversion factor between $SFR$ and $L_{2-10\,\mathrm{keV}}$ may increase with redshift, introducing a potential underestimation of star formation rates. The same conclusion is drawn by \cite{Symeonidis2011} who quantify their deviation to the \cite{Ranalli2003} calibration to be a factor of five. This would directly imply that, adopting the \cite{Symeonidis2011} conversion factor, our star formation rates would increase by a factor of five. However, for further analysis we chose to stick to the \cite{Ranalli2003} calibration just because it is the most widely used calibration throughout the literature.

\begin{figure}
\centering
\includegraphics[width=0.5\textwidth]{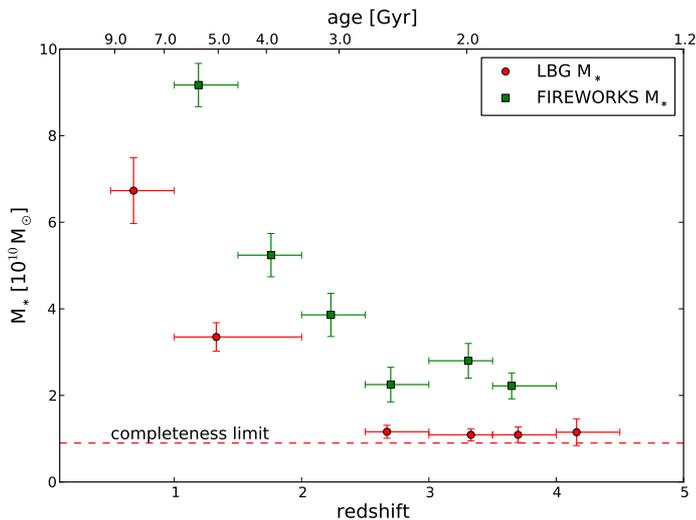}
\caption{The stellar masses of our LBG sample with redshift. For comparison, the stellar masses derived from the entire FIREWORKS sample with spectroscopic redshifts are shown, too. The plotted completeness limit corresponds to a 5$sigam$ limiting magnitude in the $K_s$-band of 22.85\,Vega-mag.}
\label{Mstar}
\end{figure}

Since our final goal is to also quantify the specific star formation rate $sSFR$, defined as the star formation rate divided by the stellar mass of the galaxy, we need to get a handle on the stellar mass. To do so, we computed the mean $K$-band magnitude of the LBGs in every stack using data from FIREWORKS \citep{Wuyts2008}. This mean $K$-band magnitude was then converted to a stellar mass following the calibration by \cite{Daddi2004}:
\begin{equation}
\log(M_*/10^{11}\,M_{\odot})=-0.4(K-K_{11})
\end{equation}
where $K$ is the total $K$-band magnitude of the respective galaxy and $K_{11}=20.15$ (Vega system) the typical $K$-magnitude of a $10^{11}\,M_{\odot}$ galaxy. A plot illustrating the results for our LBG sample is shown in Fig.~\ref{Mstar}.
Since \cite{Daddi2004} found this relationship for galaxies in the redshift range $1<z<3$, we note that our low-$z$ control sample may be affected by a systematic effects caused by its lower redshift. The scatter of this relationship is quantified by \cite{Daddi2004} to be $\sim$0.1\,dex, therefore the error on $M_*$ takes into account both this scatter as well as the 1$\sigma$ uncertainty of the mean $K$-band magnitude. The errors of the specific star formation rates are then calculated by simple Gaussian error propagation. Note that, since the derived stellar masses for our LBG sample are just above the completeness limit, the higher-redshift bins might be affected by incompleteness issues leading to an overestimation of stellar mass.

\subsection{The X-ray luminosity of  Lyman Break Galaxies}
\begin{figure}
\centering
\includegraphics[width=0.5\textwidth]{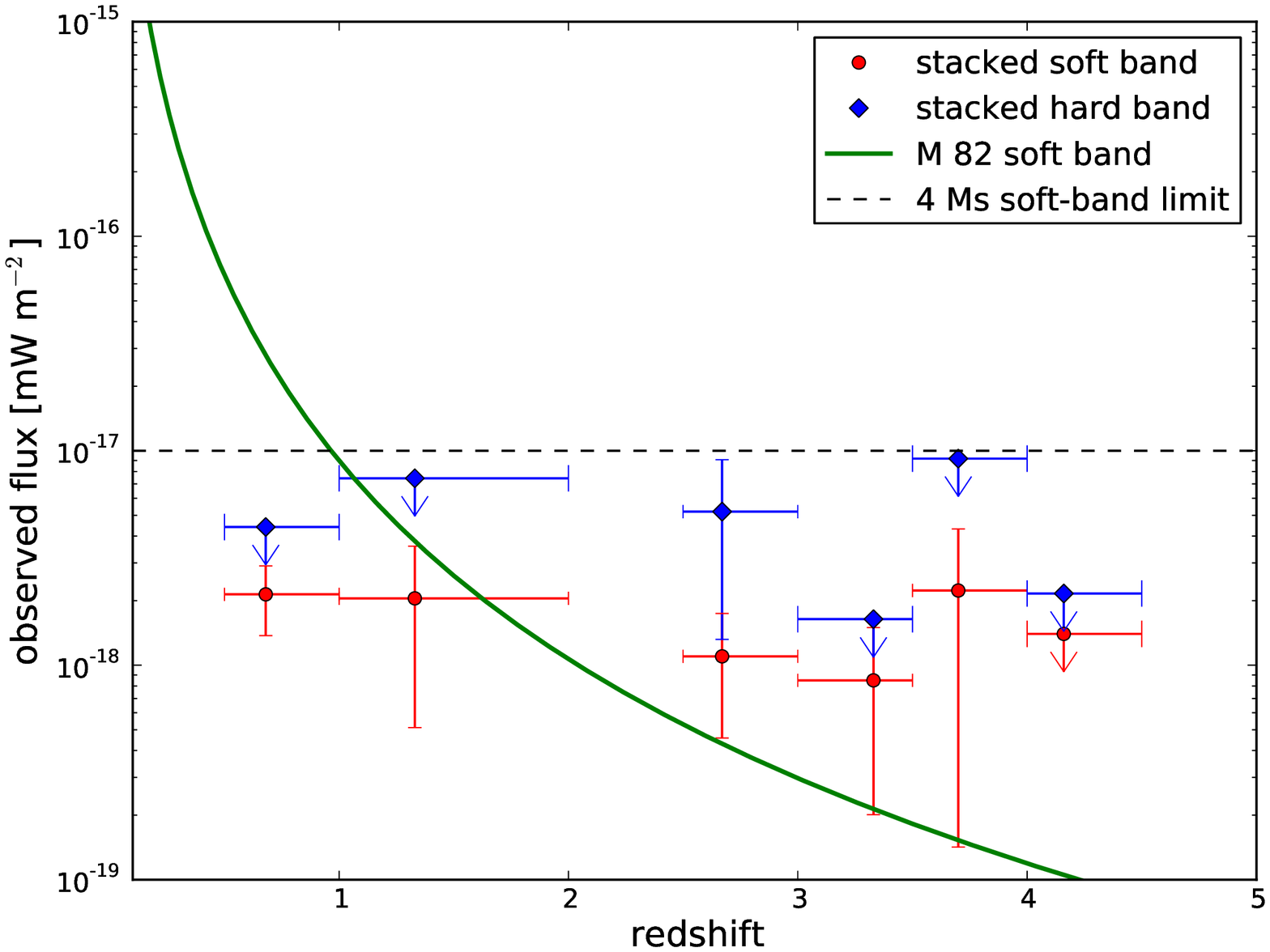}
\caption{The stacked fluxes of all samples in the observed-frame soft- and hard-band vs. redshift. Horizontal error bars indicate the width of the redshift bins, the solid line indicates the observed soft-band flux of the low-$z$ LBG analog VV\,114 from \cite{Grimes2006}. VV\,114 has an IR-derived star formation rate of $SFR=48\,M_{\odot}\,$yr$^{-1}$ \citep{Soifer1989}. The horizontal dashed line indicates the soft-band on-axis detection limit of the {\it Chandra} 4\,Ms mosaic as reported by \cite{Xue2011}.}
\label{luminosity}
\end{figure}

Fig.~\ref{luminosity} shows the measured X-ray fluxes versus redshift. In general, our results agree well with other X-ray stacking studies such as, for example, \cite{Lehmer2005} who found a rest-frame 2-8\,keV luminosity for a sample of 449 $U$-band dropouts ($z\sim3$) of $(1.5\pm0.3)\,10^{34}\,$W and $(1.4\pm0.6)\,10^{34}\,$W for a sample of 395 bright $B$ dropouts ($z\sim4$), also seeing a nearly constant luminosity with increasing redshift. A similar result was also obtained by \cite{Nandra2002} who give a 2-10\,keV rest-frame luminosity of $(3.4\pm0.7)\,10^{34}\,$W for their sample of 144 $U$-band dropouts in the {\it Hubble} Deep Field North. They also stacked a sample of 95 ``Balmer Break'' galaxies which are located at $z\sim1$, comparable to our BZ-UV sample with a median redshift of 1.33. For their $z\sim1$ sample, \cite{Nandra2002} report a 2-10\,keV rest-frame luminosity of $(0.33\pm0.05)\,10^{34}\,$W which also agrees well with our value from Tab.~\ref{params}. Furthermore, \cite{Nandra2002} only find significant detections in the soft-band stacks just as in this work. The only hard-band stacking detection is made for the Spec-U2 bin (presumably because it is the bin containing the most stacked sources) with a hard-band flux of $(5.20\pm1.88)\,10^{-18}\,$mW\,m$^{-2}$. Together with the corresponding soft-band flux (see Tab.~\ref{fluxes}), this gives a photon index $\Gamma\approx1.1$, significantly lower than the { value of 2.0 typical for star-formation activity.} Since such low photon indices are more typical for AGN, we assume that this one hard X-ray detection may be due to a small remaining contamination of an AGN which still survived the many steps of AGN rejection previously applied to all stacked sources. { An alternative explanation may be the contamination by a strong emission line at 7.47\,keV as discussed e.g. in \cite{Fiore2012}. However, since the effective area of {\it Chandra} becomes very small for such high energies, the contribution of photons with energies exceeding 7\,keV only makes up a small fraction of the total counts in the 2-8\,keV band.}

However, we highlight that recent studies by \cite{Lehmer2012} hint a steep rise of number counts of normal star-forming galaxies just below the soft-band flux limit of the {\it Chandra} 4\,Ms mosaic. They report that at flux levels of the order of $10^{-17}\,$mW\,m$^{-2}$ normal SFGs make up nearly half the X-ray number counts and that at even lower flux levels those galaxies will completely dominate the population. Therefore, since our stacking analysis probes the faint X-ray population down to $10^{-18}\,$mW\,m$^{-2}$, our findings are also consistent with extrapolations from individually detected sources.

\subsection{Star formation in Lyman Break Galaxies}
\begin{figure*}
\centering
\includegraphics[width=0.49\textwidth]{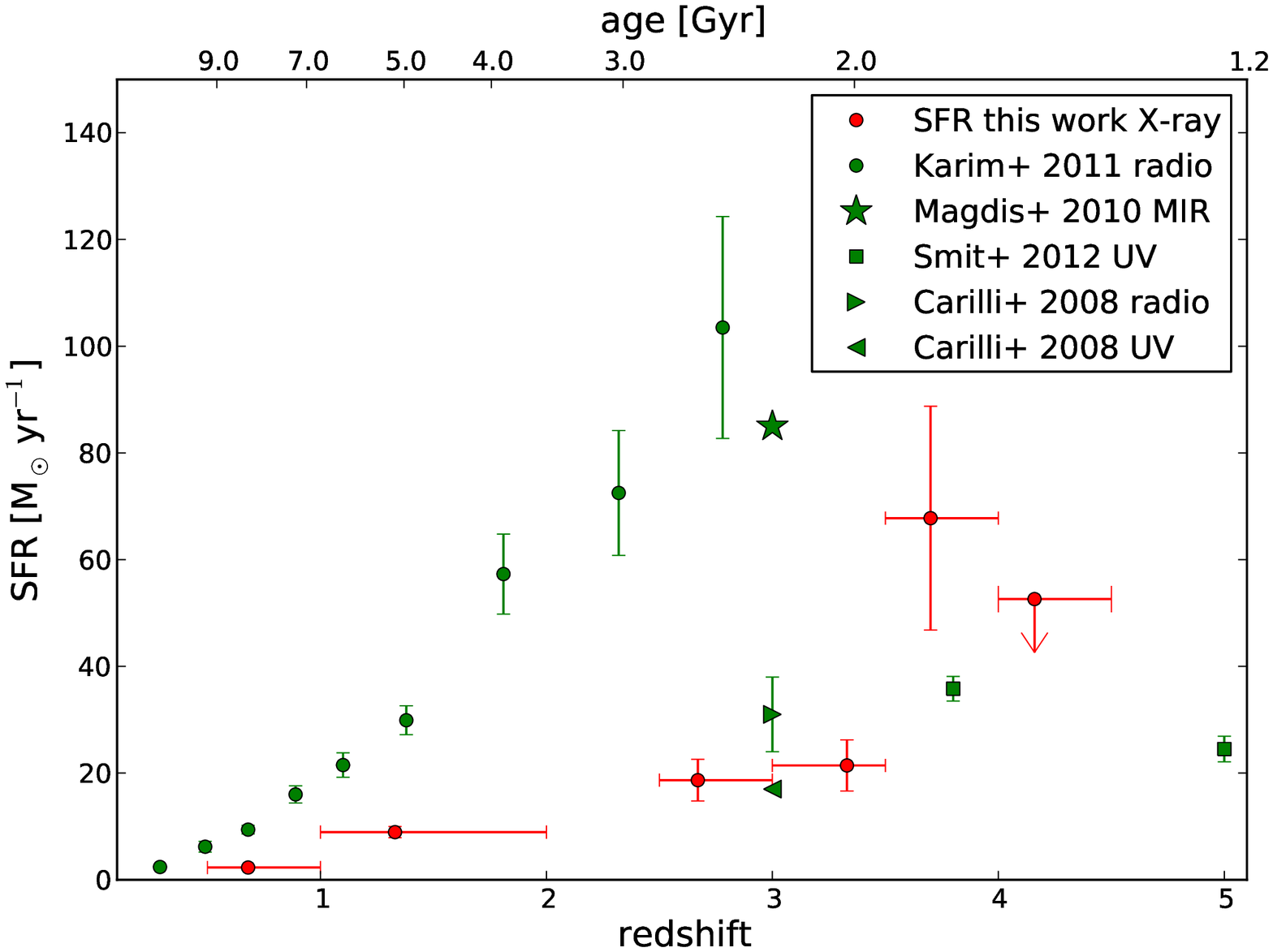}
\includegraphics[width=0.49\textwidth]{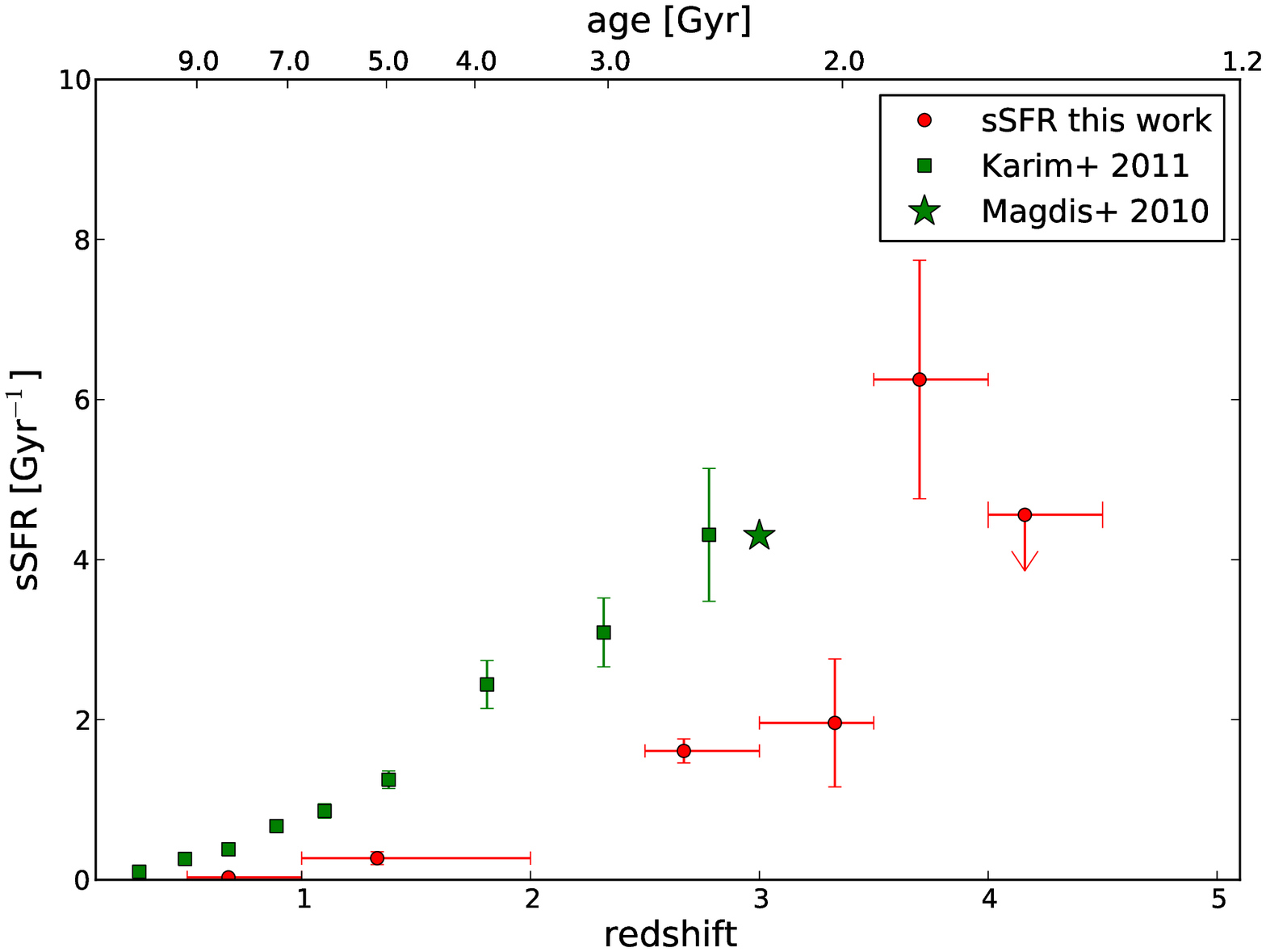}
\caption{The star formation rates (left) and the specific star formation rates (right) for different LBG samples vs. redshift. We compare our findings to the ones by \cite{Karim2011}, \cite{Magdis2010}, \cite{Carilli2008} and \cite{Smit2012} using various other star formation rate indicators such as radio or UV luminosities.}
\label{SFR}
\end{figure*}
From the rest 2-10\,keV luminosities of our various redshift bins, we estimated the star formation rates for an average Lyman Break Galaxy using the method by \cite{Ranalli2003}. The resulting SFRs are shown in Fig.~\ref{SFR} (left panel) alongside with several other SFR measurements for LBGs calculated from data taken at different wavelength ranges. Our X-ray derived SFRs peak at a redshift of about $z_{peak}=3.5$ with the last measurement at $z\sim4$ being an upper limit slightly below the peak value.

Compared to SFR estimated from other wavelength ranges, we find at least a good agreement in the trend (increase to $z_{peak}=3.5$, then decrease)whilst the absolute numbers differ by a factor 2 to 5, depending on the particular data set for comparison. Especially the \cite{Karim2011} values show much larger values which could be easily explained by the very different selection criteria they used for their stacking input samples: While we focus on LBGs, they selected their sample based on 3.6\,$\mu$m flux density to be able to split not only in redshift bins but also in stellar mass bins. However, the trend of increasing SFR to at least $z=3$ is seen in both our LBG as well as their 3.6\,$\mu$m samples. To compare with other LBG samples, we plotted the values obtained by \cite{Carilli2008} for lower redshifts and \cite{Smit2012} for higher redshifts. \cite{Carilli2008} used LBGs in the COSMOS field to create $U$-, $B$- and $V$-dropout samples as input for a 1.4\,GHz stacking analysis. They detected only the $U$-dropouts ($z\sim3$) in the radio stacks and find $SFR=31\pm7\,M_{\odot}\,\rm{yr}^{-1}$  \citep[adopting the $SFR$-$L_{\mathrm{1.4\,GHz}}$ calibration by][]{Yun2001} while the average UV-derived SFR for this sample is $17\,M_{\odot}\,\rm{yr}^{-1}$. These two values agree well with our X-ray based estimate of $21.43\pm4.79\,M_{\odot}\,\rm{yr}^{-1}$ at this redshift. The decrease with redshift is confirmed by the results of \cite{Smit2012} who investigated the star formation in higher-redshift LBGs based on their rest-frame UV continuum slopes. Accounting for dust extinction following \cite{Meurer1999}, they find SFRs of 35$\,M_{\odot}\,\rm{yr}^{-1}$ and 24$\,M_{\odot}\,\rm{yr}^{-1}$ at redshifts of 3.8 and 5.0, respectively. Comparing their results to other data, they see a peak in star formation rate between $3.0<z_{peak}<3.5$ and a peak value of 50-60$\,M_{\odot}\,\rm{yr}^{-1}$, very comparable to our X-ray based findings.

All together, we can confirm a peak in the star formation activity of LBGs at a redshift around $z_{peak}=3.5$ which is seen by various authors using various SFR estimation methods. However, the absolute values for star formation rate differ by quite a large margin which is either due to the different selection criteria for the galaxy samples investigated or inconsistencies in the calibration of the different SFR estimation methods in different wavelength ranges, see e.g. \cite{Kurczynski2010} or . This again shows that a uniform and consistent cross-calibration of SFR indicators is desperately needed, particularly in advance of upcoming large survey projects observing across the entire electromagnetic spectrum.

Regarding the specific star formation rates (sSFR), a similar peaking trend with redshift is seen (Fig.~\ref{SFR}, right panel). Although our galaxies are in general less massive than in the \cite{Karim2011} selection, the two samples are in better agreement while looking at the sSFR since it normalizes for the mass. Therefore it looks like the LBG sample is smoothly continuing the 3.6\,$\mu$m selection at higher redshifts. This suggests that at redshifts of $z\sim4$, LBGs are typical for the galaxy population whereas at lower redshifts more massive galaxies are abundant. Since the upper limit obtained for $z=4.16$ is significantly lower than the actual measurement at $z=3.70$, a clear decrease for higher redshifts is also visible in sSFR. This peaking behavior of the specific star formation rate supports the widely adopted picture of stellar mass growth having a peak somewhere between redshift 2 and 4, but contradicts findings by other authors who observed \citep{Feulner2005} or modeled \citep{Khochfar2011} the sSFR for a wide range of redshifts and find a more or less constant sSFR (at least for lower mass galaxies) from $z=2$ onwards to higher redshifts. An explanation for this difference could again be the different sample selection criteria. While we looked at LBGs only, \cite{Feulner2005} did a more complex near-IR selection to again split in stellar mass bins. A comparable LBG sample in the Subaru Deep Field was investigated by \cite{Yoshida2006} who also find a peaking sSFR with a peak value of about 0.1\,Gyr$^{-1}$ at $z\approx4$. This is more than an order of magnitude lower than our peak value, but since they focus on more massive LBGs this difference is not a surprise. They argue \citep[according to the analytical galaxy evolution model by][]{Hernquist2003} that the trend of a peaking SFR/sSFR at redshifts between 3 and 4 can be explained by different parameters dominating the star formation process at various redshifts: At lower redshifts, the star formation activity is mostly governed by the cooling rate of the (molecular) gas residing in a dark matter halo whereas at higher redshifts the conversion of cold gas into stars is the dominating parameter regulating star formation. The transition between these two modes is marked by the peak of star formation activity. This hypothesis is recently supported by the work of \cite{Reddy2012} who also see a (mild) peak in the specific star formation rate of several Gyr$^{-1}$ at $z\approx3$.

\subsection{The contribution of LBGs to the cosmic star formation rate density}
\begin{figure}
\centering
\includegraphics[width=0.5\textwidth]{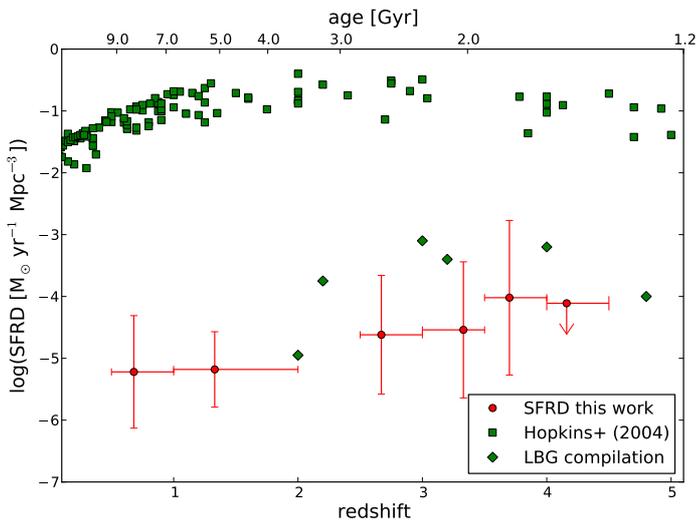}
\caption{The contribution of LBGs to the total cosmic star formation rate density (SFRD). SFRD measures (green squares) were taken from \cite{Hopkins2004}. LBG compilation (green diamonds) refers to measurements compiled by \cite{Yoshida2006} which include values from \cite{Wyder2005} and \cite{Arnouts2005}.}
\label{SFRD}
\end{figure}
Fig.~\ref{SFRD} shows the co-moving star formation rate density (SFRD) as derived from our star formation rates (Tab.~\ref{params}) and a cosmology according to Sect.~\ref{intro}. As one can easily see, the peaking trend also continues in this plot, although the peak is much less pronounced. We compare our estimates to values from \cite{Yoshida2006} and references therein (green diamonds in Fig.~\ref{SFRD}) to find a similar peaking trend albeit at absolute SFRD values being about one order of magnitude higher than ours. These higher values can be explained by the difference in LBG sample selection between \cite{Yoshida2006} and our work (all our LBGs are spectroscopically confirmed whereas they employ a mixed spectroscopic/photometric selection process).

However, despite the (relatively) small difference between the two LBG samples, there is a much larger difference when comparing SFRDs derived from LBG samples to estimates of the total cosmic star formation rate density as e.g. compiled by \cite{Hopkins2004}. The difference between total and LBG SFRD is about four orders of magnitude, hence deeming LBGs to be entirely negligible when investigating the bulk of star formation activity in the universe's history, even when considering potential incompleteness of our LBG sample. This finding agrees very well with previous investigations of the cosmic star formation history \citep[e.g.][]{Sawicki2006,Bouwens2007,Bouwens2009}. From constraints on the UV luminosity functions at various redshifts, these authors drew conclusions for the question on whether the main fraction of star formation activity takes place at the luminous or faint end of the high-redshift galaxy population. Finding extremely steep faint-end slopes at all redshifts $z>2$, they argue that most of the UV emission (and hence star formation) takes place in low-luminosity galaxies with only a few percent contribution of the bright population. More quantitatively, \cite{Sawicki2006} derived characteristic luminosities $\mathcal{L}_*$ (and the corresponding absolute magnitude $\mathcal{M}_*$) for LBGs by fitting \cite{Schechter1976} luminosity functions (LFs) at redshifts 2.2, 3.0 and 4.0. They find a nearly constant value of $\mathcal{M}_*=-21.0$ at a rest-frame wavelength of 1700\,$\mathrm{\AA}$, corresponding to a star formation rate of about $15\,M_{\odot}\,\rm{yr}^{-1}$ (uncorrected for potential dust extinction). Because of the steep faint-end slopes of the fitted LFs, they argue that the total UV luminosity density (and hence star formation rate density) in this redshift range not dominated by $\mathcal{L}_*$ or even brighter gaalxies but by faint galaxies, mostly around luminosities of $0.1\,\mathcal{L}_*$. The faint end-slopes of the rest-frame UV LFs becomes even steeper at higher redshifts \citep{Bouwens2007,Bouwens2009}.
Therefore, the marginal contribution of LBGs to the total cosmic SFRD is not surprising since the LBG samples used in this work are based on a spectroscopic selection. Hence the sources must be bright enough for spectroscopy, which in this case \citep[e.g. for the VIMOS spectroscopic campaign in the CDF-S by][]{Balestra2010} means that they have to have apparent magnitudes brighter than 24.5 in the $B$- and $R$-bands, respectively. At $z=2.0$, this corresponds to an absolute B-band (so rest-frame $\sim1500\,\mathrm{\AA}$) magnitude of -21.5, half a magnitude brighter than $\mathcal{M}_*$ derived by \cite{Sawicki2006}. At $z=3.0$, the situation becomes even worse since the spectroscopic sample in this case is limited to objects 1.5 magnitudes brighter than $\mathcal{M}_*$. We therefore point out that our results are in very good agreement with the previous statements that the bulk of star formation activity takes place not in bright but in faint galaxies. An independent study recently conducted by \cite{Tanvir2012} utilizing gamma-ray bursts (GRBs) as star formation rate tracers, so following an entirely different premise, comes to the same result. With their data for even higher redshifts ($z\sim5$), they argue that the bulk of star formation activity is not even accessible in currently available ultra-deep data sets such as the {\it Hubble} Ultra-Deep Fields which implies mean star formation rates per ``typical'' galaxy of less than $0.2\,M_{\odot}\,\rm{yr}^{-1}$.

\section{Summary and Conclusions}
Utilizing stacking techniques together with the newly acquired {\it Chandra} 4\,Ms mosaic, we have investigated the X-ray luminosity and star formation activity of Lyman Break Galaxies (LBGs) across cosmic time. Our stacking input sample spans a redshift range $0.5<z<4.5$, making use of the superb spectroscopic data in the {\it Chandra} Deep Field South region. Spectroscopic selection of LBGs guarantees a highly homogeneous sample with minimal contamination by interloping objects, in particular AGN contaminating the measured X-ray fluxes. Our newly developed stacking algorithm, optimized for highest sensitivities in exchange for the loss of morphological information allows us to probe the LBG population down to formerly unprecedented flux density levels of $10^{-18}\,$mW\,m$^{-2}$ in the soft (0.5-2\,keV) band. Our findings are summarized below:
\begin{enumerate}
\item We reliably (Poissonian confidence level $>95\%$) detect X-ray emission from our LBG input sample out to redshifts of about 4 with a robust upper limit for $z=4.5$.
\item From the observed soft-band fluxes, we derive rest-frame 2-10\,keV luminosities using a photon index $\Gamma=2.0$ and correcting for Galactic hydrogenabsorption. They show a nearly constant value with redshift, underlining the robustness of our stacking procedure.
\item From the rest 2-10\,keV luminosities we calculated mean star formation rates for each redshift bin. Our results show a distinct peak at $z_{peak}\sim3.5$, in good agreement with the general trend of various other estimates of SFR for LBGs and other types of galaxies. However, we point out that the various star formation rate indicators across the electromagnetic spectrum deliver significantly different SFR values, therefore a thorough comparison and cross-calibration is highly needed also in advance of upcoming all-sky survey projects.
\item With ancillary $K$-band infrared data, we calculated stellar masses and hence specific star formation rates (sSFR=SFR/M$_*$). The peaking behavior of our LBG sample is also seen in sSFR, underlining the  widely adopted picture of stellar mass growth having a peak somewhere between redshift 2 and 4.
\item Considering the contribution of LBGs to the total cosmic star formation rate density (SFRD), we find that LBGs only make up a tiny fraction of the total star formation activity at all investigated redshifts, supporting the emergent notion that the bulk of star formation in the universe takes place in very low-mass, low-$SFR$ galaxies currently escaping detection with all available facilities.
\end{enumerate}
All together, Lyman Break Galaxies seem to be good examples of ``typical'' galaxies at their respective redshifts but are not the place in the universe where most the star formation activity takes place. Despite their high typical star formation rates of a few to several 10$\,M_{\odot}\,\rm{yr}^{-1}$ at all redshifts, LBGs are just not abundant enough to make up a significant fraction of the total star formation in the universe. This highly supports the proposition that the bulk of high-redshift star formation is going on in faint, yet undetected galaxies well below $\mathcal{L}_*$ at all redshifts $z>2$ \citep[see e.g.][and references therein]{Tanvir2012}.

\begin{acknowledgements}
We thank our anonymous referee for his/her constructive comments, in particular regarding the many technical finesses of X-ray observations.
\end{acknowledgements}

\bibliographystyle{aa}
\bibliography{bib}

\end{document}